\documentclass{revtex4}
\usepackage{amssymb}
\usepackage{amsmath}

\input{tcilatex}

\begin{document}

\title[]{Colliding plane wave solution in $F(R)=R^{N}$ gravity }
\author{S. Habib Mazharimousavi}
\email{habib.mazhari@emu.edu.tr }
\author{M. Halilsoy}
\email{mustafa.halilsoy@emu.edu.tr}
\author{T. Tahamtan}
\email{tayabeh.tahamtan@emu.edu.tr}
\affiliation{Department of Physics, Eastern Mediterranean University, G. Magusa, north
Cyprus, Mersin 10, Turkey. }

\begin{abstract}
\textbf{Abstract:} We identify a region of $F\left( R\right) =R^{N}$ gravity
without external sources which is isometric to the spacetime of colliding
plane waves (CPW). From the derived curvature sources, $N$ ($N>1$) measures
the strength (i.e. the charge) of the source. The analogy renders
construction and collision of plane waves in $F\left( R\right) =R^{N}$
gravity possible, as in the Einstein-Maxwell (EM) theory, simply because $R=0
$. A plane wave in this type of gravity is equivalent to a Weyl curvature
plus an electromagnetic energy-momentum-like term (i.e. 'source without
source'). For $N=1$ we recover naturally the plane waves (and their
collision) in Einstein's theory. Our aim is to find the effect of an
expanding universe by virtue of $F(R)=R^{N}$ on the colliding gravitational
plane waves of Einstein.
\end{abstract}

\maketitle

\section{INTRODUCTION}

$F\left( R\right) $ gravity in which the Einstein-Hillbert action (i.e. $%
F\left( R\right) =R$) gets corrections by an arbitrary function of scalar
curvature $R$ has become much popular in recent years. Such an interest
stems from cosmological observations and attempts to interpret the
accumulated data. The accelerated expansion of our universe with dark energy
/ matter, phantom fields, quintessence and cosmological constant can all be
addressed under a common modified theory entitled as $F\left( R\right) $
gravity. Dependence on the single (and simplest) Ricci scalar $R$ becomes an
advantage when compared with theories depending on the higher order
invariants. The $F\left( R\right) $ gravity, even in the absence of external
sources is highly non-linear, but recent studies show that many exact
solutions have been obtained \cite{1} (for the review of the subject). No
doubt, the freedom of adding suitable energy momentum in addition to those
arising from curvature makes the theory further complicated, but interesting.

In this paper we wish to draw attention to another aspect of $F\left(
R\right) $ gravity which is local equivalence (isometry) with the spacetime
of colliding plane waves (CPW) (for the review of the subject see \cite{2})
. It was shown first by Chandrasekhar and Xanthopoulos (CX) in 1986 \cite{3}
that part of black hole spacetime is locally isometric with the spacetime of
CPW \cite{3}.

For black hole spacetimes with single (event) horizon, such as
Schwarzschild, the region of isometry is inside the horizon where the
geometry admits two space-like Killing vectors. For black holes that admit
double horizons (i.e. $r=r_{-}=$ the \ inner and $r=r_{+}=$ the outer
horizon) such as Reissner-Nordstr\"{o}m (RN) and Kerr, the dynamical
isometry region occurs in between the two horizons, $r_{-}<r<r_{+}$. This
property has been used in the past to generate CPW solutions from the known
solutions of black holes \cite{4}. Specifically, this technique has aided in
generating regular classes of CPW solutions due to the fact that the
singularity of the black hole is removed in the isometry process. Let us
note that the intrinsic connection between black holes and plane waves is
not limited by the observation in \cite{3}. Much earlier Penrose proved a
Theorem \cite{5} in this connection: "Every spacetime admits a plane wave
spacetime as a limit", and naturally black hole spacetimes are not exempted
from this statement. For a class of extremal black holes we have shown even
further, that the near horizon geometry coincides with that of CPW \cite{6}.
Being motivated by these past studies it is our aim to point out that
similar isometries exist also in $F\left( R\right) =R^{N}$ gravity and in a
larger class of $F\left( R\right) $ theories which admit RN solution of
Einstein's theory as a particular solution. To be on the safe side, however,
in the satisfaction of $F\left( R\right) $ equations in the limit $%
R\rightarrow 0,$ we restrict ourselves to the $F\left( R\right) =R^{N}$
case. In this study we consider a $4$ -dimensional spherically symmetric $%
F\left( R\right) $ gravity solution without external sources. That is, the
curvature creates its own source through the intrinsic nonlinearity. This
derived source, or more popularly, "source without source", turns out to be
an electromagnetic-like source and the geometry is defined modulo the RN
geometry. Under a suitable transformation this part transforms into the
spacetime of CPW. The set of boundary (junction) conditions for the CPW is
adopted from the Einstein-Maxwell (EM) theory in which the appropriate
conditions are those of O'Brien and Synge \cite{7}. These guarantee that
non-physical terms, such as the product of two Dirac delta functions, or any
other undesired combinations do not arise at all. It should be added that a
general analysis of junction conditions in more general $F\left( R\right) $
gravity other than $F\left( R\right) =R^{N}$, deserves a separate study.

By construction our solution of CPW admits a horizon following the
collision, in place of a spacetime singularity. Whether the horizon is
stable or not, and if stable / unstable, against what type of perturbations,
remain open to which we shall not address in this paper. Further, by going
from the interaction region (Region IV) to the incoming regions (Regions II
and III) we identity, in analogy with taking the Penrose-limit \cite{5}, the
incoming plane waves prior to the instant of collision. Region I is flat or
the no-wave zone. Extension into the incoming regions give us a class of
plane waves in $F\left( R\right) =R^{N}$ gravity which consist of a
gravitational (Weyl) and an energy-momentum part derived from the curvature.
In this picture ($N-1$) plays the role of the electromagnetic charge so that
we assume $N>1$.

Similar analysis can be considered for different versions of $F\left(
R\right) $ gravity, provided the function $F\left( R\right) $ is known
explicitly. Let us note that in most of the cases considered in cosmology a
closed form(i.e. analytical) representation for $F\left( R\right) $ is not
available. From physics standpoint a plane wave in general relativity is a
representation of an infinitely boosted particle under proper limiting
procedure \cite{8}. At such high speeds a particle loses its matter property
and turns into a light-like wave. For this reason in the large hadron
collision (LHC) experiments the particle can alternatively and approximately
be considered as a wave. From this token it should be added that the almost
plane wave representation of an infinitely boosted proton, due to its
internal structure (i.e. colors, gluons, quarks, etc.) is yet to be found.
Our boosting process covers so far only point-like objects without internal
structure. In this regard what is available at our disposal is no more than
a coarse approximation to the exact case. Another important item is that the
horizons formed in CPW are unstable against arbitrary perturbations \cite{9}%
. This may be the classical justification that micro black holes formed at
high (TeV scale) energies are not stable / detectable. From the quantum
picture the probability of black hole (or any other objects, such as mini
wormholes) formation is non zero. However, if these objects are to be
detected classically their stability must be verified. In this case the
stability arguments of horizons in CPW may offer some hints. It is believed
that modified theory of gravity such as $F\left( R\right) $ with its rich
content will give tangible results.

Organization of the paper is as follows. Section II summarizes $F\left(
R\right) $ gravity with its equivalence to the space of CPW. Section III
discusses the properties of our colliding wave spacetime. Comparison of
Einstein and $F(R)=R^{N}$ theories takes place in Section IV. Conclusion and
Discussion in Section V completes the paper.

\section{$F\left( R\right) =R^{N}$ gravity and local equivalence with the
space of CPW.}

We consider the action in the form ($8\pi G=1$)

\begin{equation}
S=-\frac{1}{2}\int d^{4}x\sqrt{-g}\left[ F\left( R\right) +L_{m}\right] ,
\end{equation}%
where the function $F\left( R\right) $ will be determined by the field
equations in general and $L_{m}$ stands for the matter Lagrangian. In this
paper we shall confine ourselves to the particular case in which $L_{m}=0$,
so that any 'source', $T_{\mu \nu }^{c}$ will be determined by the
curvature. Further, we shall follow an easier route to assume $F\left(
R\right) =R^{N}$ a priori, and use it to find the curvature sources. With
these assumptions Einstein's equations are still valid as 
\begin{equation}
G_{\mu \nu }=T_{\mu \nu }^{c}=\frac{1}{F^{\prime }\left( R\right) }\left[ 
\frac{1}{2}g_{\mu \nu }\left( F\left( R\right) -RF^{\prime }\left( R\right)
\right) +\left( \nabla _{\mu }\nabla _{\nu }-g_{\mu \nu }\square \right)
F^{\prime }\left( R\right) \right] ,
\end{equation}%
where $F^{\prime }\left( R\right) =\frac{dF}{dR}$ and $\square =\nabla _{\mu
}\nabla ^{\mu }$ . When the geometrical components of $G_{\mu \nu }$ are
substituted into the foregoing equation it yields 
\begin{equation}
R_{\mu \nu }F^{\prime }\left( R\right) -\nabla _{\mu }\nabla _{\nu
}F^{\prime }\left( R\right) +\left( \square F^{\prime }\left( R\right) -%
\frac{1}{2}F\left( R\right) \right) g_{\mu \nu }=0.
\end{equation}%
By taking the trace of the field equation one obtains%
\begin{equation}
F^{\prime }\left( R\right) R+3\square F^{\prime }\left( R\right) -2F\left(
R\right) =0.
\end{equation}%
Now, assuming a spherically symmetric metric ansatz%
\begin{equation}
ds^{2}=f\left( r\right) dt^{2}-\frac{dr^{2}}{f\left( r\right) }-r^{2}\left(
d\theta ^{2}+\sin ^{2}\theta \text{ }d\varphi ^{2}\right) ,
\end{equation}%
in which $f\left( r\right) $ is the only function to be found and the choice 
$F\left( R\right) =R^{N}$ yields \cite{10} 
\begin{gather}
R^{N-1}\left[ N\left( -\frac{1}{2}\frac{rf^{\prime \prime }+2f^{\prime }}{r}%
\right) -\frac{1}{2}R\right] =0, \\
R^{N-1}\left[ N\left( -\frac{rf^{\prime }+\left( f-1\right) }{r^{2}}\right) -%
\frac{1}{2}R\right] =0, \\
R^{N}\left( 1-\frac{N}{2}\right) =0, \\
\left( f^{\prime }=\frac{df}{dr}\right) .  \notag
\end{gather}%
The latter equation admits either $N=2$ or $R=0$ which is the case of our
interest here. The other equations also confine the parameter $N$ to $N>1$
in order to be satisfied. Therefore setting $R=0$ admits a solution for $%
f\left( r\right) $ as 
\begin{equation}
f\left( r\right) =1-\frac{2m}{r}+\frac{\sigma }{r^{2}}.
\end{equation}%
Here $m$ and $\sigma $ are two integration constants. Throughout the
analysis we shall make the choice $\sigma \neq 0$, since otherwise $f\left(
r\right) $ reduces to that of Schwarzschild, and also our choice will be
that of $N>1$. This form of $f\left( r\right) $ is easily identified as the
RN black hole where $m$ and $\sigma $ are the mass and square of the charge,
respectively. In the following analysis we shall make the simple choice $m=1$
without loss of generality. This class of solutions can be considered in
three different cases, depending on the horizon(s), namely 
\begin{equation}
r_{\pm }^{2}-2r_{\pm }+\sigma =0.
\end{equation}%
This admits the roots%
\begin{equation}
r_{\pm }=1\pm \alpha ,
\end{equation}%
for 
\begin{equation}
\alpha \mathbf{=}\sqrt{1-\sigma }.
\end{equation}%
We can identity three different classes:

\textbf{i) }$\alpha =1$\textbf{, or }$\sigma =0$: This brings us back to
Einstein's gravity, so it is not interesting for our purpose.

\textbf{ii) }$\alpha =0$\textbf{, or }$\sigma =1$: This choice corresponds
to the extremal RN whose near horizon geometry is the Bertotti-Robinson (BR)
spacetime. In other words, we have 
\begin{equation}
f\left( r\right) =\left( 1-\frac{1}{r}\right) ^{2}.
\end{equation}%
Now, making the transformation%
\begin{equation}
r=1+\lambda \tilde{r},\text{ \ \ \ \ \ }t=\frac{\tilde{t}}{\lambda },
\end{equation}%
in the limit, $\lambda \rightarrow 0$, with the successive replacements $%
\tilde{r}=\frac{1}{R}$ and $\tilde{t}=T$ yields for the line element 
\begin{equation}
ds^{2}=\frac{1}{R^{2}}\left( dT^{2}-dR^{2}\right) -d\theta ^{2}-\sin
^{2}\theta \text{ }d\varphi ^{2}.
\end{equation}%
This is the well-known BR spacetime of EM theory which can be transformed
easily into the space of CPW by a well-established transformation \cite{2}.
The resulting spacetime will be the Bell-Szekeres space time for colliding
electromagnetic shock waves \cite{11}. For this reason we shall not be
interested in this particular case of $\alpha =0$, either.

\textbf{iii) }$0<\alpha <1$\textbf{\ (or }$0<\sigma <1$\textbf{): }In this
paper we shall be interested in this particular class since (\textbf{i}) and
(\textbf{ii}) will not add much to our interest. From now on we can make the
choice $0<\sigma <1$ ($0<\alpha <1$), which is the consequence of choosing $%
N>1$ from Eq.s (6-8) . Our line element (5) reads now

\begin{equation}
ds^{2}=\left( 1-\frac{2}{r}+\frac{\sigma }{r^{2}}\right) dt^{2}-\frac{dr^{2}%
}{\left( 1-\frac{2}{r}+\frac{\sigma }{r^{2}}\right) }-r^{2}\left( d\theta
^{2}+\sin ^{2}\theta \text{ }d\varphi ^{2}\right) .
\end{equation}%
In between the horizons

\begin{equation}
1-\alpha =r_{-}<r<r_{+}=1+\alpha ,
\end{equation}%
$t$ becomes space-like and $r$ becomes time-like. Now let us apply the
transformation

\begin{equation}
r=1+\alpha \tau ,\text{ \ }\cos \theta =\sigma ,\text{ \ }\varphi =y,\text{
\ }t=x,
\end{equation}%
to the new coordinates $\left( \tau ,\sigma ,x,y\right) $ from the old ones $%
\left( t,r,\theta ,\varphi \right) $. Next, we transform once more from $%
\left( \tau ,\sigma \right) $ to the null-coordinates $\left( u,v\right) $ by

\begin{equation}
\tau =\sin \left( a\text{ }u+b\text{ }v\right) ,\text{ \ \ \ \ }\sigma =\sin
\left( a\text{ }u-b\text{ }v\right) ,
\end{equation}%
in which $\left( a,b\right) $ are constants that we introduce for
convenience. These measure the magnitudes of incoming energies and in case
that we choose $a=1$ ($b=1$) we normalize those strengths to unity. After an
overall scaling of coordinates $x$, $y$ and $ds^{2}$ by the constant $2ab$
the line element takes the form

\begin{equation}
ds^{2}=\left( 1+\alpha \text{ }\tau \right) ^{2}\left( 2du\text{ }dv-\delta
dy^{2}\right) -\frac{\Delta }{\left( 1+\alpha \text{ }\tau \right) ^{2}}%
dx^{2},
\end{equation}%
in which we have abbreviated $\Delta =1-\tau ^{2}$ and $\delta =1-\sigma
^{2} $. So far, we have only applied coordinate transformations on our
original metric (5) until we reach at (20). It is a remarkable observation
at this stage that the metric (20) can be interpreted as a CPW metric under
the substitutions

\begin{equation}
u\rightarrow u\text{ }\theta \left( u\right) ,\text{ \ \ \ \ }v\rightarrow v%
\text{ }\theta \left( v\right) ,
\end{equation}%
where $\theta \left( u\right) $ and $\theta \left( v\right) $ are the
Heaviside step functions corresponding to the two different null boundaries $%
u=0$ and $v=0$. The adopted O'Brien-Synge \cite{7} junction conditions
guarantee in this process that no extra sources arise at the null
boundaries. Otherwise the metric can not be interpreted as a CPW spacetime
and the idea should be discarded.

\section{Properties of the colliding wave spacetime}

Upon insertion of the step functions our CPW reads

\begin{eqnarray}
ds^{2} &=&\left( 1+\alpha \text{ }\sin \left[ au\text{ }\theta \left(
u\right) +bv\text{ }\theta \left( v\right) \right] \right) ^{2}\left( 2du%
\text{ }dv-\cos ^{2}\left[ au\text{ }\theta \left( u\right) -bv\text{ }%
\theta \left( v\right) \right] dy^{2}\right) -  \notag \\
&&\frac{\cos ^{2}\left[ au\text{ }\theta \left( u\right) +bv\text{ }\theta
\left( v\right) \right] }{\left( 1+\alpha \text{ }\sin \left[ au\text{ }%
\theta \left( u\right) +bv\text{ }\theta \left( v\right) \right] \right) ^{2}%
}dx^{2}.
\end{eqnarray}%
With the choice of null-basis $1-$forms of Newman and Penrose (NP) (in which
the step functions present)

\begin{eqnarray}
l &=&\left( 1+\alpha \tau \right) du,  \notag \\
n &=&\left( 1+\alpha \tau \right) dv, \\
m &=&\sqrt{\frac{\delta }{2}}\left( 1+\alpha \tau \right) dy+i\sqrt{\frac{%
\Delta }{2}}\frac{dx}{\left( 1+\alpha \tau \right) }  \notag
\end{eqnarray}%
all non-vanishing spin-coefficients and NP quantities are tabulated in
Appendix A. Fig. 1 depicts the details of the space time of CPW. The Region
I $\left( u<0\text{, }v<0\right) $ is flat. The Region II $\left(
u>0,v<0\right) $has the wave profile

\begin{equation}
\Psi _{4}\left( u\right) =-\alpha ^{2}\delta \left( u\right) +\frac{3\alpha
a^{2}\theta \left( u\right) \left( \alpha +\sin \left( a\text{ }u\right)
\right) }{\left( 1+\alpha \sin \left( a\text{ }u\right) \right) ^{4}},
\end{equation}%
which is the superposition of an impulsive and a shock gravitational wave.
Since our class of solutions is valid for $0<\alpha <1$ the impulse wave has
trailing shock waves and both components of impulsive and shock waves are
indispensable. The only surviving energy-momentum component in Region II is

\begin{equation}
\Phi _{22}\left( u\right) =\frac{a^{2}\theta \left( u\right) \left( 1-\alpha
^{2}\right) }{\left( 1+\alpha \sin \left( a\text{ }u\right) \right) ^{4}}.
\end{equation}%
The incoming Region III has the same structure as Region II with $%
a\rightarrow b$, $u\rightarrow v$ so that the non-vanishing components are $%
\Psi _{0}\left( v\right) $ and $\Phi _{00}\left( v\right) $ (see Appendix
A3). After collision, in the interaction region (Region IV) $\left( u>0\text{%
, }v>0\right) $ we have the additional components of $\Psi _{2}\left(
u,v\right) $\ and $\Phi _{02}\left( u,v\right) =\Phi _{20}\left( u,v\right) $
which didn't exist in the incoming regions.

It is observed from the NP invariants that the interaction region $\left( u>0%
\text{, }v>0\right) $ has no singularities (i.e. $A\left( u,v\right) =\left(
1+\alpha \text{ }\sin \left( au\text{ }+bv\text{ }\right) \right) \neq 0$.
On the null-boundaries, however, (i.e. $au=\frac{\pi }{2},v=0$ and $bv=\frac{%
\pi }{2},u=0$) there are null-singularities since the Weyl components
diverge there. In the interaction region $\left( u>0\text{, }v>0\right) $
none of the curvature invariants diverge, the metric function $g_{xx}$ ,
however, vanishes which indicates a coordinate singularity. It can be
recalled that before applying the coordinate transformation $g_{xx}$ was $%
g_{tt}$ whose zero gave rise to the horizon. The hypersurface $\sin \left( au%
\text{ }+bv\text{ }\right) =1$, is the horizon that forms as a result of
colliding waves. From the CPW experience in the EM theory it is known that
such horizons are unstable against perturbations \cite{9}. That is, addition
of any small physical field gives rise to a back reaction effect which makes
the horizon to transform into a true spacetime singularity. By extension
beyond the horizon (i.e. $au+bv>\frac{\pi }{2}$) it is evident, for example
somewhere within $\frac{\pi }{2}<au+bv<\frac{3\pi }{2},$ that we face a
curvature singularity. It is expected that the same behavior takes place
also in $F\left( R\right) $ gravity. \ Our example of solution is simple
enough to simulate the CPW in EM theory. Any solution that will be found in
this regard must inherit a more violent wave interaction and stronger
singularity. In the interaction region $\left( u>0\text{, }v>0\right) $ we
observe that off-the null boundaries the Dirac delta functions vanish and
only the shock terms survive. Under this condition the Weyl curvature
components satisfy

\begin{equation}
9\Psi _{2}^{2}=\Psi _{0}\Psi _{4},
\end{equation}%
which implies that the spacetime is of type-D, expectedly since after all it
is a transform of a type-D spacetime. In this spacetime it can easily be
shown that the massive scalar wave equation and Hamilton-Jacobi equations
can easily be solved by separation of variables. Let us add that the null
coordinates $\left( u,v\right) $ is not an appropriate coordinate system for
this purpose. A more appropriate coordinate system is given by $\left( \tau
,\sigma ,x,y\right) $, so that the scalar field equation is

\begin{equation}
\left( \square +m\right) \Phi \left( \tau ,\sigma ,x,y\right) =0,
\end{equation}%
where $\square $ is the covariant Laplacian and $m$ is the mass of the
particle. We seek a separability in the form

\begin{equation}
\Phi \left( \tau ,\sigma ,x,y\right) =T\left( \tau \right) S\left( \sigma
\right) e^{i\mu _{0}x+iv_{0}y},
\end{equation}%
where $\mu _{0}$ and $v_{0}$ are constants and the functions $T\left( \tau
\right) $ and $S\left( \sigma \right) $ are functions of their arguments.
The decoupled equations are reduced into quadratures

\begin{gather}
\left( \delta S_{\sigma }\right) _{\sigma }=k_{0}S,  \notag \\
\left( \Delta T_{\tau }\right) _{\tau }+\left( mA^{2}-k_{0}\right) T=0.
\end{gather}%
with the separation constant $k_{0}$ , $A\left( \tau \right) $ is defined in
the Appendix A2) and sub $\left( \tau \right) $ / $\left( \sigma \right) $
indices imply partial derivatives. In the next Section we shall make a
comparison between general relativity and $F(R)$ gravity results in both the
NP quantities and the restricted geodesics.

\section{Comparison between $F(R)=R$ and $F(R)=R^{N}$ ($N>1$)}

\subsection{The Ricci and Weyl tetrad scalars}

All the related tetrad scalars are given in the Appendix (A3-A5). For $%
\alpha =1$ they reduce to $F(R)=R$ which are the results of general
relativity. Now, in order to compare with the case of $F(R)=R^{N}$ we set 
\begin{equation}
\alpha =1-\epsilon 
\end{equation}%
in all quantities to the first order with $\epsilon >0$ and $\epsilon
^{2}\simeq 0$. Upon substitution and expansion in first order of $\epsilon $
we obtain the following forms 
\begin{eqnarray}
\Phi _{ij} &=&\Phi _{ij}^{\left( 0\right) }+\epsilon \Phi _{ij}^{\left(
1\right) }+\mathcal{O}\left( \epsilon ^{2}\right)  \\
\Psi _{i} &=&\Psi _{i}^{\left( 0\right) }+\epsilon \Psi _{i}^{\left(
1\right) }+\mathcal{O}\left( \epsilon ^{2}\right) 
\end{eqnarray}%
in which $\Phi _{ij}^{\left( 0\right) }$ and $\Psi _{i}^{\left( 0\right) }$
are the Ricci and Weyl components in the Einsteins general relativity. The $%
\Phi _{ij}^{\left( 1\right) }$ and $\Psi _{i}^{\left( 1\right) }$ are
naturally the first order contributions from the $F(R)$ gravity. As a result
the expressions (A3-A5) are obtained as follow%
\begin{eqnarray}
\Phi _{00}^{\left( 0\right) } &=&\Phi _{22}^{\left( 0\right) }=\Phi
_{02}^{\left( 0\right) }=0,\Psi _{2}^{\left( 0\right) }=\frac{ab}{\Sigma ^{4}%
},  \notag \\
\Psi _{4}^{\left( 0\right) } &=&\frac{-a\delta \left( u\right) }{\cos
bv\left( 1+\sin bv\right) ^{2}}+\frac{3a^{2}}{\Sigma ^{3}},\Psi _{0}^{\left(
0\right) }=\frac{-b\delta \left( v\right) }{\cos au\left( 1+\sin au\right)
^{2}}+\frac{3b^{2}}{\Sigma ^{3}},  \notag \\
\Phi _{00}^{\left( 1\right) } &=&\frac{2b^{2}}{\Sigma ^{4}},\Phi
_{22}^{\left( 1\right) }=\frac{2a^{2}}{\Sigma ^{4}},\Phi _{02}^{\left(
1\right) }=\frac{2ab}{\Sigma ^{4}},\Psi _{2}^{\left( 1\right) }=\frac{%
ab\left( 3\sin K-2\right) }{\cos bv\left( 1+\sin bv\right) ^{2}}, \\
\Psi _{4}^{\left( 1\right) } &=&\frac{-a\delta \left( u\right) \left( 3\sin
bv-1\right) }{\cos bv\left( 1+\sin bv\right) ^{3}}+\frac{3a^{2}\left( 3\sin
K-2\right) }{\Sigma ^{4}},  \notag \\
\Psi _{0}^{\left( 1\right) } &=&\frac{-b\delta \left( v\right) \left( 3\sin
au-1\right) }{\cos au\left( 1+\sin au\right) ^{3}}+\frac{3a^{2}\left( 3\sin
K-2\right) }{\Sigma ^{4}}.  \notag
\end{eqnarray}%
in which $\Sigma =1+\sin K$ and $K$ is defined in Appendix A. We have also%
\begin{eqnarray}
Z^{\left( 0\right) }\left( u\right)  &=&\frac{2-\sin au}{\cos au},Z^{\left(
1\right) }\left( u\right) =\frac{2}{\cos au}\left( \frac{\sin au-1}{\sin au+1%
}\right) , \\
Z^{\left( 0\right) }\left( v\right)  &=&\frac{2-\sin bv}{\cos bv},Z^{\left(
1\right) }\left( v\right) =\frac{2}{\cos bv}\left( \frac{\sin bv-1}{\sin bv+1%
}\right) .  \notag
\end{eqnarray}%
By choosing $u<0$/$v<0$ in the above expressions we find the differences
between the incoming plane waves between the two theories. This may be
interpreted as the distortion of Einstein's plane gravitational waves in the
expanding universe model due to the $F(R)=R^{N}$ gravity. These expressions
complete the amplitudes of $F(R)=R^{N}$ gravity in comparison with the
general relativity at the first order.

\subsection{Deviation of Geodesics}

We consider the geodesics projected into the ($u,v$) plane (i.e. $x=$cons., $%
y=$const.) and for simplicity we fix the constants $a=b=1$. The system can
be described by the Lagrangian%
\begin{equation}
L\left( u,v\right) =\left( 1+\alpha \sin \left( u+v\right) \right) \sqrt{%
v^{\prime }}
\end{equation}%
in which $v^{\prime }=\frac{dv}{du},$ so that $v$ is parametrized in terms
of $u$. With this much simplification the geodesics equation follows as%
\begin{equation}
v^{\prime \prime }=\frac{2\alpha v^{\prime }\left( 1-2v^{\prime }\right)
\cos \left( u+v\right) }{1+\alpha \sin \left( u+v\right) }.
\end{equation}%
Obviously an exact solution follows for the choice $v=\frac{1}{2}u,$ but in
general the differential equation is highly nonlinear which can be expressed
in terms of elliptic functions. Since our aim is to compare geodesics with
general relativity we take again $\alpha =1-\epsilon $, as we did above and
make expansions in terms of $\epsilon $. We obtain to the first order in $%
\epsilon $ the following differential equation%
\begin{equation}
v^{\prime \prime }=2v^{\prime }\left( 1-2v^{\prime }\right) \frac{\cos
\left( u+v\right) }{1+\sin \left( u+v\right) }\left( 1-\frac{\epsilon }{%
1+\sin \left( u+v\right) }\right) .
\end{equation}%
Fig. (2) displays the numerical plots of Einstein's ($\epsilon =0$) and
geodesics for the small parameters $\epsilon =0.2.$ The initial speeds $%
v^{\prime }=\frac{dv}{du}$ is varied accordingly. The geodesics take place
within the region $u+v<\frac{\pi }{2},$ restricted by the horizon shown by
the broken line in Fig. (2).

\section{Conclusion and discussion}

The pathological problems associated with the singularities and horizons in
colliding waves were exclusively discussed in the past (see \cite{9} and
references cited therein) so that we didn't feel the necessity of repeating
them here. Our aim is to draw attention to a well-known local isometry
between CPW and black holes that exist also in the nowadays fashionable $%
F\left( R\right) $ gravity. We have chosen the particular case of $F\left(
R\right) =R^{N}$, ($N>1$) as an example in which the scalar curvature $R=0$.
Our solution is single parametric with parameter $\alpha \mathbf{=}\sqrt{%
1-\sigma }$ . The employed solution is locally isometric to a portion of the
Reissner-Nordstr\"{o}m (RN) geometry and effectively our $F\left( R\right)
=R^{N}$ gravity becomes Einstein-Maxwell (EM)-like. The physical
interpretation, however, is entirely different. There are no separate
Maxwell equations for instance. In this way from the local isometry we
construct CPW solutions from a particular $F\left( R\right) $ gravity. By
interpolating into the incoming regions we obtain a class of plane waves
(i.e. the Penrose limit) in this theory. In general, the Weyl curvature of
these plane waves are in the form of (Impulsive) + (Shock) waves. Its
energy-momentum in all regions consists of pure shock terms, as it should
be. From the EM analogy the boundary (junction) conditions that has been
adopted are those of O'Brien- Synge which eliminate any spurious boundary
currents. We remark that for more general (and further nonlinear) structure
of $F\left( R\right) $ theories new boundary conditions are required. We
admit that our metric of CPW is the simplest one in which by virtue of $R=0$
all pertinent $F\left( R\right) $ equations are automatically satisfied. We
find all deviations of the Weyl and Ricci scalars from the Einstein's
gravity and compare also the geodesics to the first order. In analogy with
the colliding electromagnetic \cite{11} and Yang-Mills \cite{13} shock waves
our interaction region here also forms a horizon through which the spacetime
can be extended analytically \cite{14}. As the next step we may consider the
case of $R=R_{0}=$constant. This will give rise to a cosmological constant 
\cite{12} whose colliding wave interpretation can not be considered free of
null shells \cite{15}. Finally we wish to comment on the importance of the
collision problem presented in this paper. Infinitely boosted particles can
be represented by plane waves in general relativity. Head-on collision of
such plane waves simulates high energy particle collisions. For this reason
production and detection of exotic objects such as microscopic-black holes /
wormholes becomes a reality provided their classical stability analysis is
verified. No doubt the richer structure of $F\left( R\right) $ gravity and
the mathematical analysis of horizon stability incorporated in CPW can be
helpful in this context.

a) The non-zero spin coefficients are (for $0<\alpha <1$)

\begin{eqnarray}
\epsilon &=&\frac{\alpha \text{ }b\theta \left( v\right) \cos K}{2A^{2}} 
\TCItag{A1} \\
\gamma &=&-\frac{\alpha \text{ }a\theta \left( u\right) \cos K}{2A^{2}} 
\notag \\
\mu &=&-\frac{a\theta \left( u\right) }{2A}\left( \tan K+\tan L\right) 
\notag \\
\rho &=&\frac{b\theta \left( v\right) }{2A}\left( \tan K-\tan L\right) 
\notag \\
\lambda &=&\frac{\text{ }a\theta \left( u\right) }{2A}\left( Z-\tan L\right)
\notag \\
\sigma &=&-\frac{b\theta \left( v\right) }{2A}\left( Z+\tan L\right)  \notag
\end{eqnarray}%
where we have used the abbreviations

\begin{eqnarray}
A &=&1+\alpha \sin K,  \TCItag{A2} \\
K &=&a\text{ }u\text{ }\theta \left( u\right) +b\text{ }v\text{ }\theta
\left( v\right) ,  \notag \\
L &=&a\text{ }u\text{ }\theta \left( u\right) -b\text{ }v\text{ }\theta
\left( v\right) ,  \notag \\
Z &=&\frac{\sin K+\alpha \left( 1+\cos ^{2}K\right) }{A\cos K}.  \notag
\end{eqnarray}

\bigskip b) The non-zero Ricci components are

\begin{eqnarray}
\Phi _{00}\left( u,v\right) &=&\frac{b^{2}\theta \left( v\right) \left(
1-\alpha ^{2}\right) }{A^{4}},  \TCItag{A3} \\
\Phi _{22}\left( u,v\right) &=&\frac{a^{2}\theta \left( u\right) \left(
1-\alpha ^{2}\right) }{A^{4}},  \notag \\
\Phi _{20}\left( u,v\right) &=&\Phi _{02}\left( u,v\right) =\frac{ab\theta
\left( u\right) \theta \left( v\right) \left( 1-\alpha ^{2}\right) }{A^{4}}.
\notag
\end{eqnarray}

c) The non-zero Weyl components are%
\begin{eqnarray}
\Psi _{2}\left( u,v\right) &=&\frac{\alpha ab\theta \left( u\right) \theta
\left( v\right) \left( \alpha +\sin K\right) }{A^{4}},  \TCItag{A4} \\
\Psi _{0}\left( u,v\right) &=&-\frac{b\delta \left( v\right) }{2A^{2}\left(
u\right) }\left( Z\left( u\right) +\tan \left( au\theta \left( u\right)
\right) \right) +\frac{3\alpha b^{2}\theta \left( v\right) \left( \alpha
+\sin K\right) }{A^{4}},  \notag \\
\Psi _{4}\left( u,v\right) &=&-\frac{a\delta \left( u\right) }{2A^{2}\left(
v\right) }\left( Z\left( v\right) +\tan \left( bv\theta \left( v\right)
\right) \right) +\frac{3\alpha a^{2}\theta \left( u\right) \left( \alpha
+\sin K\right) }{A^{4}},  \notag
\end{eqnarray}%
in which $\delta \left( u\right) $ and $\delta \left( v\right) $ stand for
the Dirac delta functions and $Z\left( u\right) $ and $Z\left( v\right) $
are 
\begin{gather}
Z\left( u\right) =\frac{\sin \left( au\theta \left( u\right) \right) +\alpha
\left( 1+\cos ^{2}\left( au\theta \left( u\right) \right) \right) }{\cos
\left( au\theta \left( u\right) \right) \left( 1+\alpha \sin \left( au\theta
\left( u\right) \right) \right) },  \tag{A5} \\
Z\left( v\right) =Z\left( u\rightarrow v\right) ,  \notag \\
\left( Z\left( 0\right) =2\alpha \right) .  \notag
\end{gather}

\textbf{Figure caption}

\textbf{Figure 1:} The $\left( u,v\right) $ plane of CPW consists of four
disjoint regions: Region I $\left( u<0,v<0\right) $ is flat. Regions II $%
\left( u>0,v<0\right) $ and III $\left( u<0,v>0\right) $ are the non-flat
incoming regions and Region IV $\left( u>0,v>0\right) $ is highly curved
Interaction region. From the right (Region II) gravitational wave $\Psi
_{4}\left( u\right) $ (accompanied with the energy components $\Phi
_{22}\left( u\right) $) collides with $\Psi _{0}\left( v\right) $ from the
left (superposed with the energy term $\Phi _{00}\left( v\right) $) at $%
u=0=v $. The collision develops the Region IV (the Interaction Region $%
u>0,v>0$) in which the nonzero curvature (and energy) components are shown.
The hypersurface $au+bv=\frac{\pi }{2}$ is a horizon whose intersections
with the null boundaries (i.e. $au=\frac{\pi }{2},v=0$ and $bv=\frac{\pi }{2}%
,u=0$ shown with dark points) are null singularities. The lines (i.e. $au=%
\frac{\pi }{2},v<0$ and $bv=\frac{\pi }{2},u<0$) are horizons of the
incoming regions.

\textbf{Figure 2: }Particular geodesics in the $(u,v)$ plane of colliding
waves according to the differential equation (35). Curves labelled by A are
the geodesics of general relativity while B's are the $F(R)$ geodesics when
the infinitesimal parameter is taken as $\epsilon =0.2.$ Depending on the
initial speed $v^{\prime }=\frac{dv}{du},$ the difference of two classes is
revealing. In particular, for $v^{\prime }>1$ the difference becomes
significant while approaching the horizon given by $u+v=\frac{\pi }{2}.$ The
only geodesic that remains invariant in going from $F(R)=R$ to $F(R)=R^{N}$
is the case $v^{\prime }=0.5$ which is evident from Eq. (36) in the text. In
effect, the expansion of the universe due to $F(R)$ gravity manifests itself
in the deviation of geodesics as displayed here.

\bigskip

\end{document}